\documentstyle[12pt]{article}

\newfont{\bsy}{cmbsy10 scaled\magstep1}
\newfont{\bmi}{cmmib10 scaled\magstep1}
\newfont{\Lbmi}{cmmib10 scaled\magstep3}
\newcommand{\bga}{\mbox{\bmi \char'013}}

\newcommand{\bgve}{\mbox{\bmi \char'042}}
\newcommand{\bgs}{\mbox{\bmi \char'033}}

\begin{document}

\begin{flushright}
NUB 3114\\
March 1995
\end{flushright}
\pagestyle{empty} \vspace{1in}

\begin{center}
{\Large {\bf Fermions in an External  $\hspace{-6pt} \mbox{ \Lbmi \char'123
\char'125}{\bf (2)}$ Magnetic Field}}\\
\vspace{0.5in} {\sc M. H. Friedman, Y. Srivastava and A. Widom}\\
\vspace{0.3in}

{\it Physics Department\\ Northeastern University\\ Boston, MA 02115, USA\\
and\\ Dipartimento di Fisica \& INFN\\ Universita di Perugia\\ Perugia, Italy
\\ \vspace{1.0in} ABSTRACT}\\ \bigskip
\parbox{5.2in}
{We consider Fermions in a constant and uniform external $SU(2)$ magnetic
field. We find that the results for the energy levels depend on the choice
of guage potential. Choosing a Landau type guage potential yields his
results. On the other hand in another guage potential, one obtains a
different continuous eigenvalue spectrum.}
\end{center}

\newpage
\pagestyle{plain} \pagenumbering{arabic}

\section{\bf Introduction.}

A constant and uniform magnetic field for a non-abelian gauge theory
\cite{Y&M} may be produced by at least two inequivalent guage potentials.
This choice is not available for the abelian case. As we shall see
below, these two different situations lead to different results for
the energy spectrum of a Fermion placed in such a field.

There has been much reseach done on closely related topics. In reference
\cite{others} we cite only that work which impacts strongly on the
material presented in this paper.

\section{\bf The Two Guage Potentials.}

We confine our attention to $SU(2)$. Let the external magnetic field be in
the $z$ direction and in the isotopic $3$ direction, {\it i.e.}, $B_3^3$. It
is constant and uniform. Now,

$$
F_{12}^3=\partial _1A_2^3-\partial _2A_1^3+g\epsilon ^{3bc}A_1^bA_2^c\eqno (1)
$$

\subsection{Gauge Potential I:}

(let $B_3^3=B$). One choice of the gauge potential can be $A_2^3=Bx$
with all other $A_\mu ^a=0$. Then,

$$
F_{12}^3=\partial _1A_2^3=B. \eqno (2)
$$

This is the well known Landau gauge potential \cite{Landau} but for a
non-Abelian theory.

\subsection{Gauge Potential II:}

The second choice is $A_1^1=\sqrt{\frac Bg}$ and $A_2^2=\sqrt{\frac Bg}$
with all other $A_\mu ^a=0$. Then,

$$
F_{12}^3=g\epsilon ^{312}\left( \sqrt{\frac Bg}\right) \left( \sqrt{\frac Bg}
\right) =B\eqno (3)
$$
We note that the magnetic field is the same for both gauge potentials.

\section{\bf The Two Sets of Solutions to the Dirac Equation.}

Let the Fermion have isotopic and ordainary spin $\frac 12$ . The Dirac
equation \cite{Dirac} is then given by,

$$
\left( \bga \cdot ({\bf p}-g\frac{\tau _a}2{\bf A}^a)+m\beta \right) \psi
=E\psi \eqno (4)
$$
for both gauge potentials.

\subsection{Choice I:}

In this case Eq.(4) becomes,

$$
\left( \bga \cdot ({\bf p}-g\frac{\tau _3}2Bx \bgve_y)+m\beta
\right) \psi =E\psi \eqno (5)
$$
where $\bgve_y$ is a unit vector in the $y$ direction. Let
$l=gB/2$ and

$$
\psi =\Omega (x)\exp \{i(p_yy+p_zz)\}. \eqno (6)
$$
Then $\Omega (x)$ satisfies,

$$
\left( {\alpha }_xp_x-\tau _3\alpha _ylx+\alpha _yp_y+\alpha _zp_z+m\beta
\right) \Omega =E\Omega \eqno (7)
$$
where $p_x$ is a q-number, while $p_y$ and $p_z$ are c-numbers.

Let $\tau _3\Omega =t\Omega $, where $t=\pm 1$. Then Eq.(7) becomes

$$
\left( \alpha _xp_x-\alpha _ytlx+\alpha _yp_y+\alpha _zp_z+m\beta \right)
\Omega =E\Omega .\eqno (8)
$$

If we temporarily suppress the dependence on the iso-spinor, we may
express $\Omega $ in terms of two component spinors.  Thus,

$$
\Omega =\left( {\
\begin{array}{c}
\phi (x) \\
\chi (x)
\end{array}
}\right) \eqno (9)
$$

Inserting Eq.(9) into (8) and eliminating $\chi $ yields the following
equation for $\phi $,

$$
\left( \sigma _x p_x-\sigma _y tlx+\sigma _y p_y+\sigma _z p_z\right) ^2\phi
=\left( E^2-m^2\right) \phi .\eqno (10)
$$
Here, the $\sigma _i$ are the usual Pauli matricies.

We square out the left hand side of Eq.(10)and let $\sigma_z \phi =s\phi $,
where $s=\pm 1$, while using $t^2=1$, to obtain

$$
\left( p_x^2+l^2x^2+p_y^2+p_z^2-tsl-2tlp_yx\right) \phi =\left(
E^2-m^2\right) \phi .\eqno (11)
$$

We now set

$$
\phi =\left( {\
\begin{array}{c}
u(x) \\
v(x)
\end{array}
}\right) .\eqno (12)
$$

For $s=1$,

$$
u=\lambda ho_n\left( x-\frac{tp_y}l\right)
$$

while

$$
v=0,\eqno (13)
$$
whereas, for $s=-1$, $u$ and $v$ change roles. $ho_n$ is the usual
$n^{\underline{th}}$ order solution to the harmonic oscillator problem with
frequency $\omega =l$\thinspace $/m=gB/2m$. $\lambda$ is the two component
isotopic spinor satisfying $\tau_3 \lambda=t \lambda$, which we had
previously suppressed. Hence, the energy spectrum is given by

$$
E=\pm \sqrt{m^2+p_z^2+(2n+1-ts)\frac{gB}2} \eqno (14)
$$
where $n=0,1,2\cdot \cdot \cdot \infty $ as usual. These are the well known
Landau levels, but for a non-abelian theory \cite{Landau}

\subsection{Choice II:}

In this case Eq.(4) becomes

$$
\left( \bga \cdot {\bf p}-h \alpha _x \tau _1 -h \alpha _y \tau_2 + m \beta
 \right) \psi =E\psi \eqno (15)
$$

where $h=\sqrt{gB}/2$. We may now set

$$
\psi =u\exp (i{\bf p\cdot x}) \eqno (16)
$$
so that all the components of {\bf p} in Eq.(15) are c-numbers. Let

$$
u=\left( {\
\begin{array}{c}
\phi \\
\chi
\end{array}
}\right) .\eqno (17)
$$

Here, $\phi $ and $\chi $ are two component isotopic spin vectors crossed
into two component spinors to yield four component objects. Substituting
Eq.(17) into (16) into (15) gives two coupled equations for $\phi $ and
$\chi $.

$$
\left( \bgs \cdot {\bf p} -h \sigma_x \tau_1 -h \sigma_y\tau_2 \right) \phi
=\left( E+m \right) \chi \eqno (18)
$$

$$
\left( \bgs \cdot {\bf p} -h \sigma _x \tau _1 -h \sigma _y \tau_2 \right) \chi
=\left( E-m\right) \phi \eqno (19)
$$

Eliminating $\chi $ between them and simplifying the result yields,

$$
\left( {\bf p}^2+2h^2-2hp_x\tau _1-2hp_y\tau _2-2h^2\sigma _z\tau _3\right)
\phi =\left( E^2-m^2\right) \phi \eqno (20)
$$

Let $\sigma _z\phi =s\phi $ where $s=\pm 1$. We now set

$$
\phi =\left( {\
\begin{array}{c}
v \\
w
\end{array}
}\right) \eqno (21)
$$
where $v$ and $w$ are the isospin up and down components of $\phi $. We note
that $\sigma _zv=sv$ and $\sigma _zw=sw$, with $s$ being the same for these
two cases. Inserting Eq.(21) into (20) and eliminating $v$ we obtain

$$
\left[ {\cal E}^2-{\bf p}^2-2h^2(1+s)\right] \left[ {\cal E}^2-{\bf p}%
^2-2h^2(1-s)\right] w=4h^2(p_x^2+p_y^2)w\eqno (22)
$$
where ${\cal E}^2{\cal =}E^2-m^2.$ We thus find that ${\cal E}$ satisfies

$$
\left( {\cal E}^2-{\bf p}^2\right) ^2-4h^2{\cal E}^2+4h^2p_z^2=0. \eqno (23)
$$

Solving for $E$ gives,

$$
E=\pm \sqrt{{\bf p}^2+m^2+\frac{gB}2+\lambda \sqrt{gB\left( p_x^2+p_y^2+
\frac{gB}4\right) }}\eqno (24)
$$
where $\lambda =\pm 1$. It is probably easiest to write the solution for $u$
using the Pauli notation. In this case we need three sets of $2\times 2$
Pauli matricies. Thus, if $a,b,c,$and $d$ are two-component spinors, we use
the notation

$$
{\sigma _x\left( {\
\begin{array}{c}
a \\
b \\
c \\
d
\end{array}
}\right) =\left( {\
\begin{array}{c}
\sigma _xa \\
\sigma _xb \\
\sigma _xc \\
\sigma _xd
\end{array}
}\right) }\hspace{.5in}{\rho _1\left( {\
\begin{array}{c}
a \\
b \\
c \\
d
\end{array}
}\right) =\left( {\
\begin{array}{c}
c \\
d \\
a \\
b
\end{array}
}\right) }\hspace{.5in}{\tau _1\left( {\
\begin{array}{c}
a \\
b \\
c \\
d
\end{array}
}\right) =\left( {\
\begin{array}{c}
b \\
a \\
d \\
c
\end{array}
}\right) }
$$
with corresponding results for the other matricies. Thus,

$$
u=N\left[ 1+\rho _1\frac{(\bgs \cdot {\bf p}- \frac{\sqrt{gB}}2\left[
\sigma _x\tau _1+\sigma _y\tau _2\right] )}{(E+m)}\right] \left[ 1+\frac{
\tau _1(p_x+ip_y)\sqrt{gB}}{(\frac{gB}2s-\lambda \sqrt{gB\left( p_x^2+p_y^2+
\frac{gB}4\right) }}\right] \left( {\
\begin{array}{c}
v \\
0 \\
0 \\
0
\end{array}
}\right)
$$

$$
\eqno (25)
$$
where $\sigma _z v=sv$ with $v$ being a two component spinor.

\section{Conclusions}

For gauge potential I the energy spectrum is given by Eq.(14), while for gauge
potential II it is given by Eq.(24). Thus, gauge potential I yields a
discrete spectrum while gauge potential II yields a continuous
spectrum for the motion in the $x$-$y$ plane, even though the $SU(2)$
magnetic field is the same in both cases. A partial check on these
results is obtained by looking at the lowest energy levels in each
case. Thus, in Eq.(14) we set $n=0$ and $ts=1$, while in Eq.(24) we set $
\lambda =-1$ and $p_x=p_y=0$. We then have

$$
E_I=E_{II}=\sqrt{p_z^2+m^2}\eqno (26)
$$

If we go up to the next level we set $n=0$ and $ts=-1$ in Eq.(14), while in
Eq.(24) we set $\lambda =+1$ and $p_x=p_y=0$. We then have

$$
E_I=E_{II}=\sqrt{p_z^2+m^2+gB}\eqno (27)
$$

However, the energy spectrum associated with the higher levels involving
motion in the $x$-$y$ plane are very different.

We do note that the $SU(2)$ currents giving rise to the external magnetic
field in the two gauge potentials is also very different. We thus
conclude that the motion of a Fermion in an $SU(2)$ external magnetic
field depends upon the sources giving rise to the field rather than
the field itself.

\newcommand{\RMP}[3]{{\em Rev. Mod. Phys.} {\bf #1}, #2 (19#3)}
\newcommand{\Rep}[3]{{\em Phys. Rep.} {\bf #1}, #2 (19#3)}
\newcommand{\Ann}[3]{{\em Annals of Phys.} {\bf #1}, #2 (19#3)}
\newcommand{\NS}[3]{{\em Nucl. Sci.} {\bf #1}, #2 (19#3)}
\newcommand{\PR}[3]{{\em Phys. Rev.} {\bf #1}, #2 (19#3)}
\newcommand{\PRL}[3]{{\em Phys. Rev. Letts.} {\bf #1}, #2 (19#3)}
\newcommand{\PL}[3]{{\em Phys. Letts.} {\bf #1}, #2 (19#3)}
\newcommand{\NPB}[3]{{\em Nucl. Phys. B} {\bf #1}, #2 (19#3)}


\begin{thebibliography}{99}
\bibitem{Y&M} C. N. Yang and R. L. Mills, \newblock \PR{96}{191}{54}
\bibitem{others} K. Wilson,\newblock \PR {D10}{2445}{74};
G. K. Savvidy, \newblock \PL{B71}{133}{77};
H. B. Nielsen and P. Olesen, \newblock \NPB{134}{376}{78};
M. L{\"u}scher, \newblock \NPB{219}{233}{83};
F. Palumbo, \newblock \PL{B149}{143}{84};
L. Cosmai and G. Preparata, \newblock \PRL{57}{2613}{86}.
\bibitem{Landau} L.D. Landau and E. M. Lifshitz, \newblock
{\em Quantum Mechanics, Relativistic Theory}, Permagon Press, Oxford (1965)
\bibitem{Dirac} P. A. M. Dirac, \newblock {\em Proc. Roy. Soc. (London)},
{\bf A117}, 610 (1928); {\em ibid.}, {\bf A118}, 351 (1928); P. A. M. Dirac,
\newblock "The Principles of Quantum Mechanics," 4th ed., Oxford University
Press, London, 1958.
\end{thebibliography}
\end{document}